\documentclass{edp-conf}
\usepackage{graphicx}
\begin{document}

\TitreGlobal{SF2A 2005}

\title{Elemental abundances vs. kinematics in the Milky Way's disk}
\author{Girard, P.}\address{Observatoire de Bordeaux, BP 89, 33270 Floirac, France}
\author{Soubiran, C.$^1$}
\runningtitle{Abundance trends}
\setcounter{page}{237}
\index{Girard, P.}
\index{Soubiran, C.}

\maketitle
\begin{abstract}
We present the results of our investigation of three samples kinematically representative 
of the thin and thick disks and the Hercules stream using the catalogue of 
Soubiran \& Girard (2005).
We have observed abundance trends and age distribution of each component.
Our results show that the two disks are chemically well separated, they
overlap greatly in metallicity and both show parallel decreasing trends
of alpha elements with increasing metallicity, in the interval -0.80 $<$ 
[Fe/H] $<$ -0.30.
The thick disk is clearly older than the thin disk with a tentative evidence of an AMR over 2-3 Gyr 
and a hiatus in star formation before the formation of the thin disk. 
In order to improve the statistics on the disk's abundance trends, 
we have developed an automatic code, TGMET{\Large$\alpha$}, to determine 
(Teff, logg, [Fe/H], [$\alpha$/Fe]) for thousands of stellar spectra 
available in spectroscopic archives. We have assessed the performances 
of the algorithm for 350 spectra of stars bieng part of the abundance catalogue.

\end{abstract}
\section{Abundances}

We have compiled a large catalogue of stars from several studies from 
the literature presenting determinations of O, Na, Mg, Al, Si, Ca, Ti, Fe, Ni
abundances (Soubiran \& Girard 2005).
Because authors of the different studies do not use the same scales and methods in their
spectral analyses, systematics between their results have been investigated
(for more details on the construction of the catalogue see Soubiran \& Girard 2005).
The final catalogue of abundances includes 743 stars.

In order to study kinematical groups of the Milky Way's disk we need velocities and 
orbits of stars from the abundance catalogue.
We first cross-correlated the catalogue with Hipparcos (ESA 1997), selecting stars with $\pi >$10 mas 
and $\sigma_{\pi} \over \pi$ $<$0.10.
Then we have searched for radial velocities in several sources. 
Distances, proper motions and radial velocities have been combined to compute the 3 components (U, V, W) 
of the spatial velocities with respect to the Sun and the orbital parameters for 639 stars.
In addition the derivation of ages was kindly done by Fr\'ed\'eric Pont making use of the Bayesian method 
of Pont \& Eyer (2004). 
Good estimations of ages were obtained for 322 stars from the abundance catalogue. 

\section{Kinematical classification}

In order to investigate the abundance trends in the thin and the thick disks separately, 
we have classified the stars into the 2 populations, using kinematical information. 
A third component, the Hercules stream recently revisited by Famaey et al. (2004) has kinematical 
parameters intermediate between the thin disk and the thick disk. 
Its stars could have polluted previous samples of 
thick disk stars selected on kinematical criteria and thus must be taken into account.
According to the known velocity ellipsoid of these 3 populations, we assign a membership probability to each star 
and select respectively 428, 84 and 44 stars having a high probability to belong to the thin disk, 
the thick disk and the Hercules stream respectively.
The (U, V) plane of the whole sample is shown in the Fig. 8 in Soubiran \& Girard (2005).

\section{Abundance trends and ages: results and discussion}

\begin{figure*}[]
\centering
\includegraphics[width=6cm]{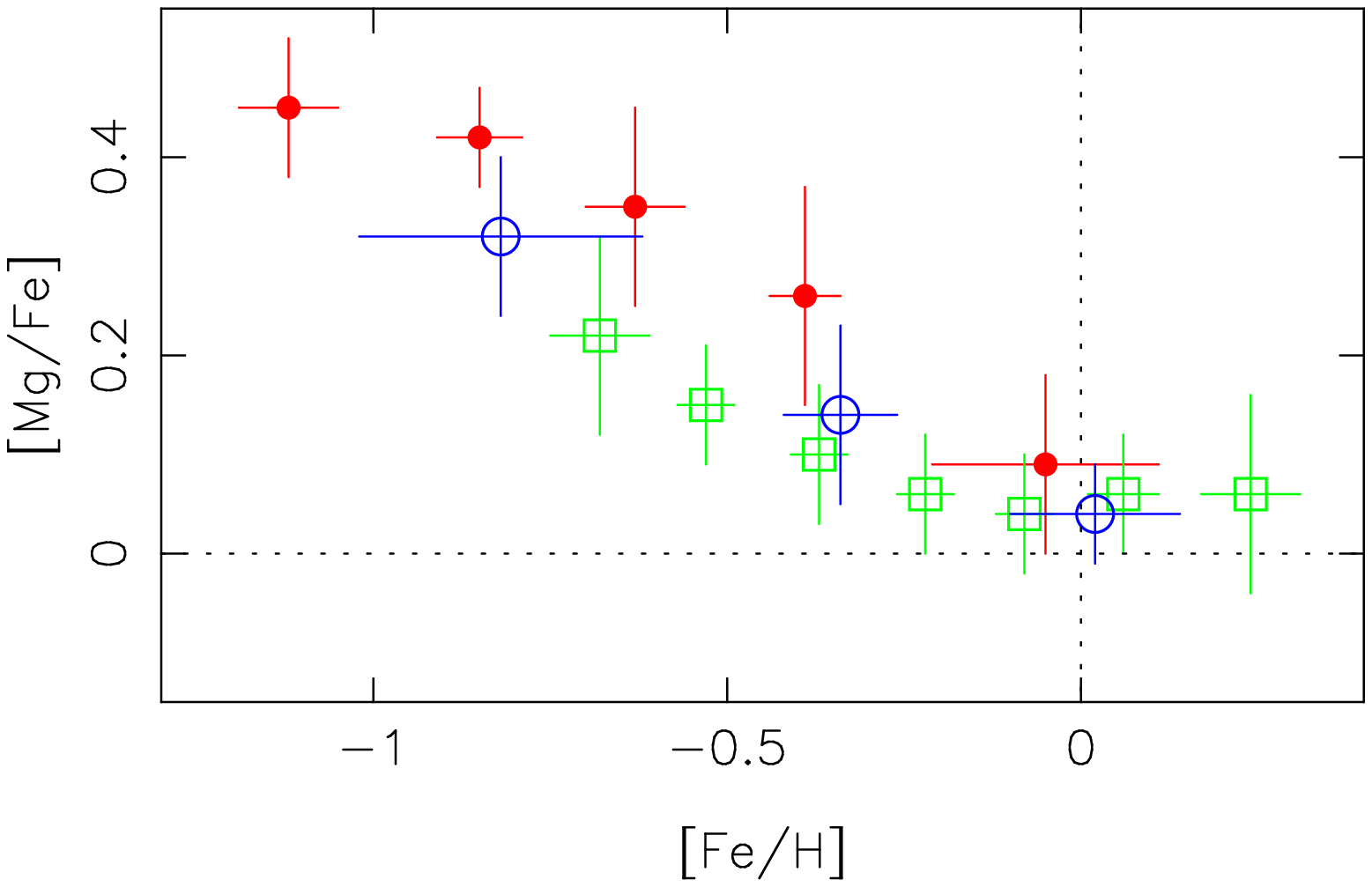}
\includegraphics[width=6cm]{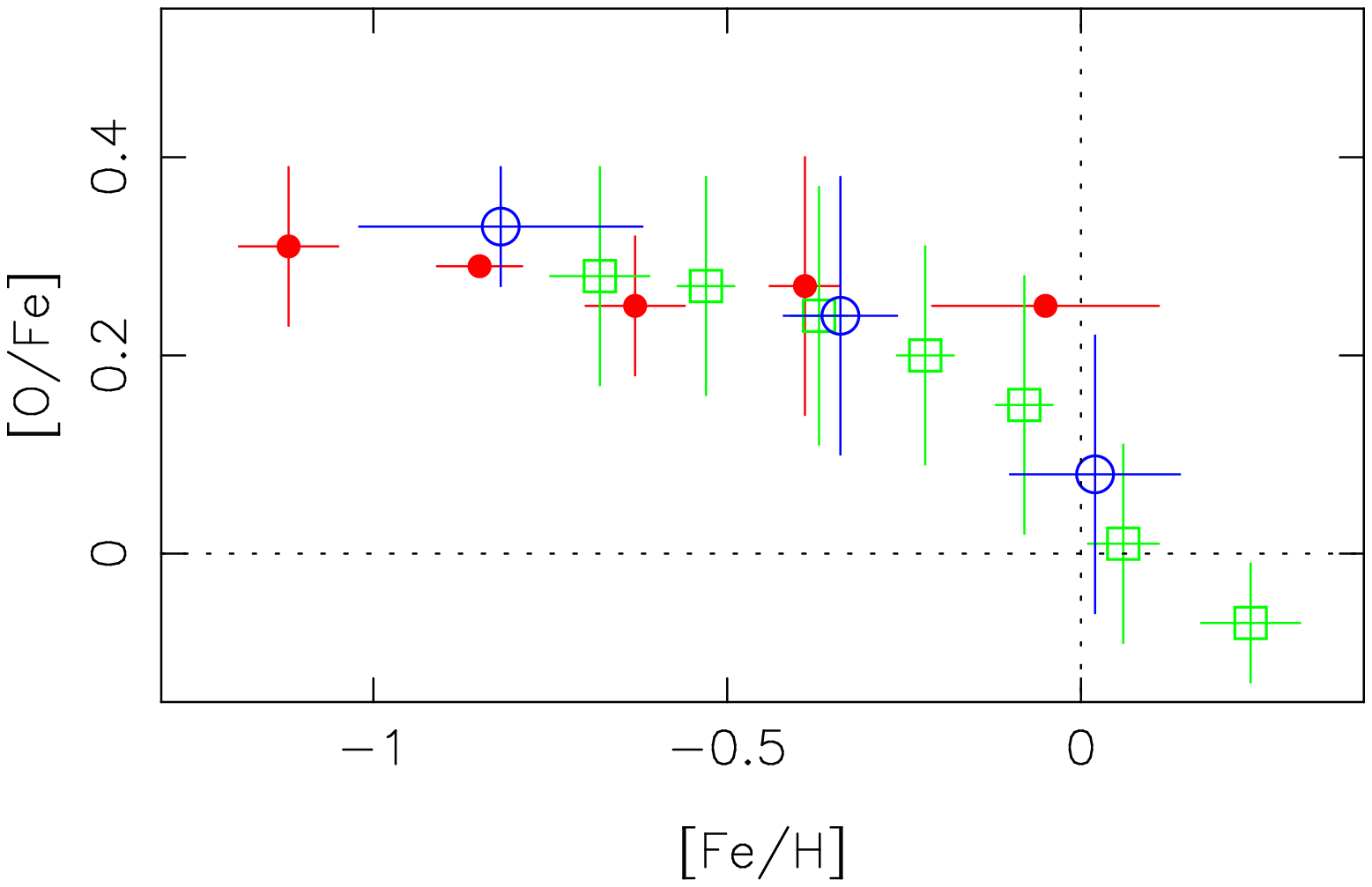}
\includegraphics[width=6cm]{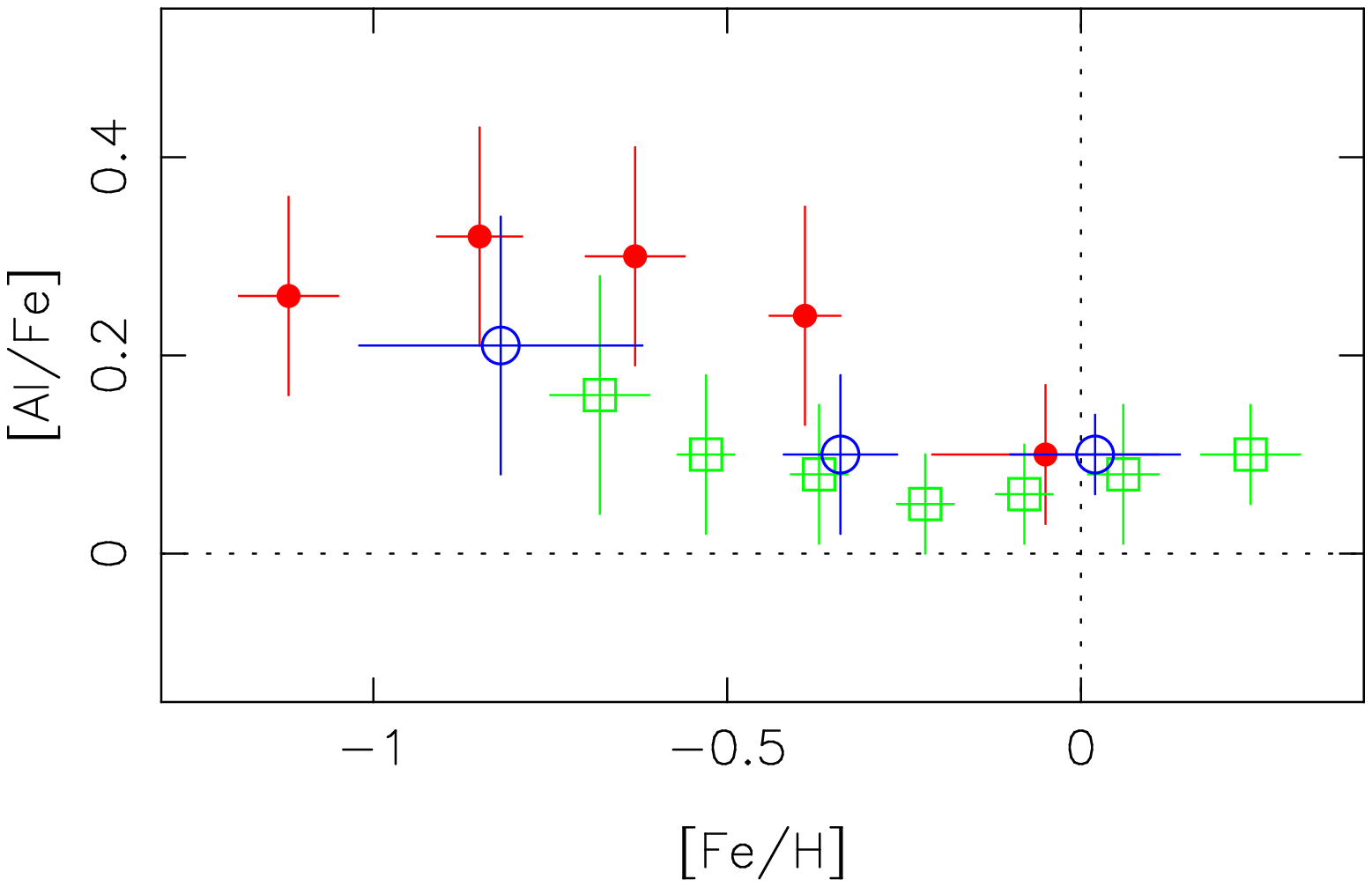}
\includegraphics[width=6cm]{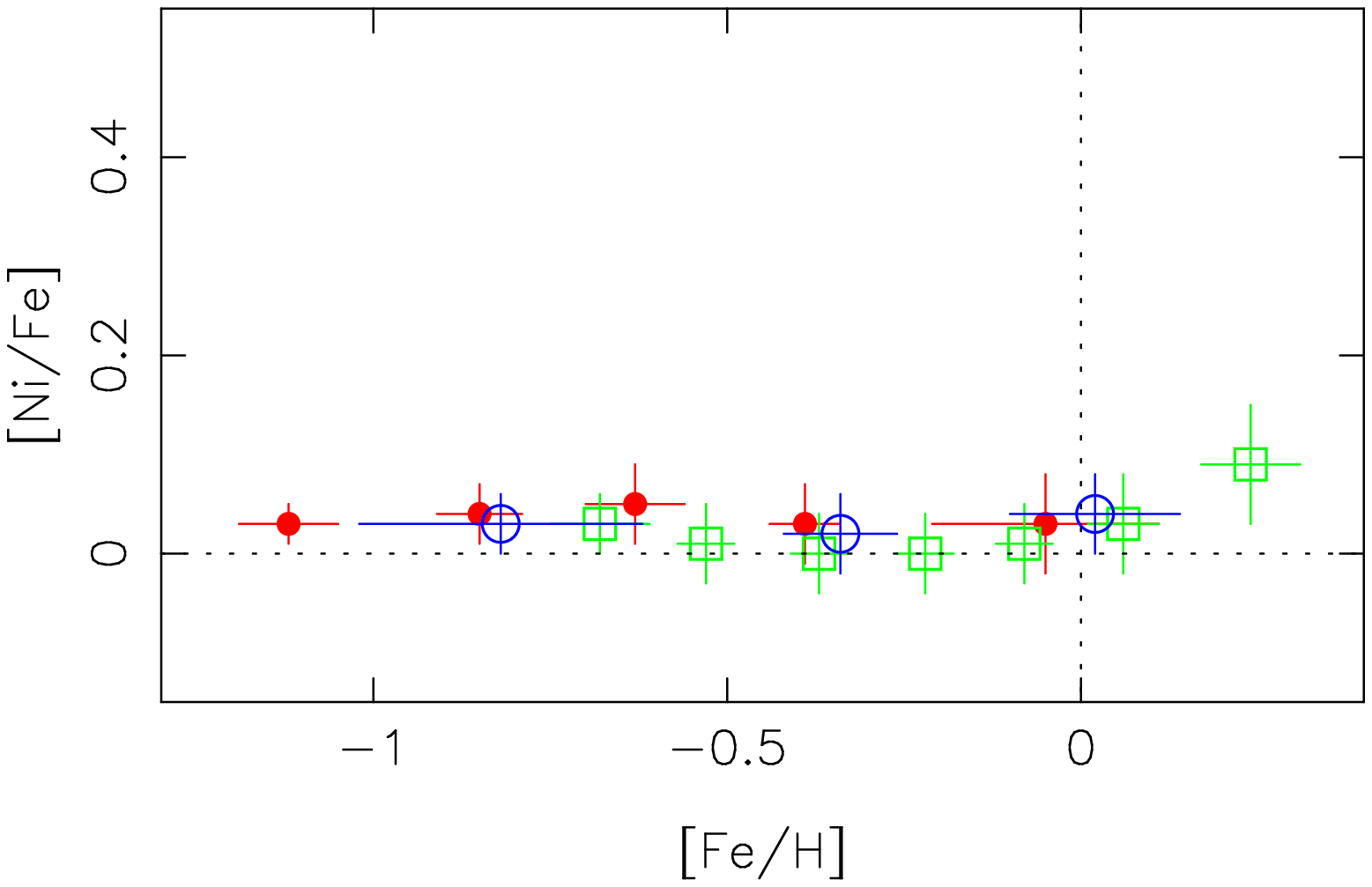}
\caption[]{Averaged [X/Fe] vs [Fe/H] per bin of metallicity in the thin disk (green squares)
in the thick disk (red filled circles) and in the Hercules stream (blue open circles).
Errors bars correspond to the standard deviations around the mean value in 
each bin. Note that Si, Ca, Ti and Na elements are not represented here but can be found in 
Soubiran \& Girard (2005).}
\label{f:figure_abu}
\end{figure*}

The abundance trends for each kinematical group are shown in Fig. \ref{f:figure_abu}.
In addition we have represented the distribution of ages vs. [Fe/H] for stars having well-defined ages for
 the thin disk, the thick disk and the Hercules stream (see Fig. 12 in Soubiran \& Girard 2005).\\ 
Our results confirm previous well established findings:
\begin{itemize}
\item The thin disk and the thick disk overlap in metallicity and exhibit parallel
slopes of [$\alpha$/Fe] vs [Fe/H] in the range -0.80 $<$ [Fe/H] $<$ -0.30, the thick disk being
enhanced.
\item The thick disk is older than the thick disk.
\end{itemize}
We bring new constraints on more controversial issues:
\begin{itemize}
\item The thin disk extends down to [Fe/H]=-0.80 and exhibits low dispersions in its abundance trends.
\item The thick disk also shows smooth abundance trends with low dispersions. The change
of slope
which reflects the contribution of the different supernovae to the ISM enrichment is visible
in [Si/Fe] vs [Fe/H] and [Ca/Fe] vs [Fe/H] at [Fe/H] $\simeq$ -0.70, less clearly in
[Mg/Fe] vs [Fe/H].
\item Al behaves as an $\alpha$ element.
\item $\lbrack$O/Fe$\rbrack$ decreases in the whole metallicity range with a change of
slope at [Fe/H] = -0.50 for the 3 populations.
\item An AMR is visible in the thin disk, the most metal-poor stars having 6.2
Gyr on average, those with solar metallicity 3.9 Gyr.
\item  Ages in the thick disk range from
7 to 13 Gyr with an average of 9.6 $\pm$ 0.3 Gyr. There is a tentative evidence of an AMR
extending over 2-3 Gyr.
\item  The most metal rich stars assigned
to the thin disk do not follow its global trends. They are significantly enhanced in all 
elements (particularly in Na and Ni) except
in O which is clearly depleted. They have also a larger dispersion in age. Half of these stars 
are probable members of the Hyades-Pleiades supercluster, two others are surprisingly 
old.
\end{itemize}

\begin{figure}[h]
   \centering
   \includegraphics[width=8.0cm , angle=90]{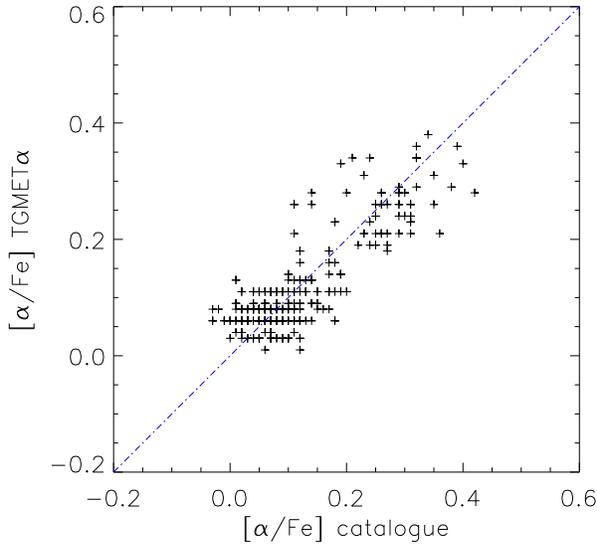}
      \caption{The [$\alpha$/Fe] ratio from TGMET{\Large$\alpha$} vs. the same ratio from 
the reference catalogue of abundances.}
       \label{figure_alf}
   \end{figure}

\section{Perspectives}
The investigation of abundances trends in the disk is currently limited by the low number 
of stars having known abundances, all being in the close solar neibourhood. In order to improve 
the statistics and reach larger distances, automatic methods of spectral analysis have to be developed. 
This is especially crucial in the perspective of GAIA which will produce millions of stellar spectra 
with substancial information on elemental abundances. We have developped a code, TGMET{\Large$\alpha$}, to determine 
automatically (Teff, logg, [Fe/H] and [$\alpha$/Fe]) on a criterion of minimum distance with respect to the 
grid of synthetic spectra of Barbuy et al. (2003). We have assessed the performances of this algorithm on high 
resolution spectra of 350 stars being part of the abundance catalogue. We compare in Fig. 2 the [$\alpha$/Fe] ratio 
obtained with TGMET{\Large$\alpha$} to those from the catalogue, considered as reference values. The low dispersion, 
0.05 dex, and the lack of systematic difference ensure that indeed reliable abundance ratios can be 
obtained automatically, at least at high resolution. Future investigation will concern low S/N or 
low resolution spectra and other grids of synthetic spectra.


\end{document}